\documentclass[twocolumn]{aastex631}
\usepackage{lineno}

\shorttitle{Abundances in the TDE ASASSN-14li}
\shortauthors{J. M. Miller}

\begin{document}

\title{Evidence of a Massive Stellar Disruption in the X-ray Spectrum of ASASSN-14li}

\correspondingauthor{Jon M. Miller}
\author[0000-0003-2869-7682]{Jon M. Miller}
\email{jonmm@umich.edu}
\affiliation{Department of Astronomy, University of Michigan, 1085 South University Avenue, Ann Arbor, MI, 48109, USA}

\author[0000-0001-6350-8168]{Brenna~Mockler}
\affiliation{Department of Physics \& Astronomy, University of
  California, Los Angeles, California, 90095}
\affiliation{The Observatories of the Carnegie Institution for
  Science, 813 Santa Barbara Street, Pasadena, California, 91101}
  
\author[0000-0003-2558-3102]{Enrico~Ramirez-Ruiz}
\affiliation{Department of Astronomy and Astrophysics,
University of California, Santa Cruz, California, 95064, USA}

\author[0000-0002-2218-2306]{Paul~A.~Draghis}
\affiliation{Department of Astronomy, University of Michigan, 1085 South University Avenue, Ann Arbor, MI, 48109, USA}

\author[0000-0002-0210-2276]{Jeremy~J.~Drake}
\affiliation{Harvard-Smithsonian Center for Astrophysics, 60
  Garden Street, Cambridge, MA 02138, USA}

\author[0000-0002-7868-1622]{John~Raymond}
\affiliation{Harvard-Smithsonian Center for Astrophysics, 60
  Garden Street, Cambridge, MA 02138, USA}

\author[0000-0003-1621-9392]{Mark~T.~Reynolds}
\affiliation{Department of Astronomy, University of Michigan, 1085 South University Avenue, Ann Arbor, MI, 48109, USA}
\affiliation{Department of Astronomy, Ohio State University,  140 W 18th Avenue, Columbus, OH, 43210, USA}

\author{Xin Xiang}
\affiliation{Department of Astronomy, University of Michigan, 1085 South University Avenue, Ann Arbor, MI, 48109, USA}

\author{Sol~Bin~Yun}
\affiliation{Department of Astronomy, University of Michigan, 1085 South University Avenue, Ann Arbor, MI, 48109, USA}

\author[0000-0002-0572-9613]{Abderahmen~Zoghbi}
\affiliation{Department of Astronomy, University of Maryland, College Park, MD, 20742, USA}
\affiliation{HEASARC, Code 6601, NASA/GSFC, Greenbelt, MD, 20771, USA}
\affiliation{CRESST II, NASA Goddard Space Flight Center, Greenbelt, MD, 20771, USA}

\begin{abstract}
The proximity and duration of the tidal disruption event (TDE)
ASASSN-14li led to the discovery of narrow, blue-shifted absorption
lines in X-rays and UV.  The gas seen in X-ray absorption is
consistent with bound material close to the apocenter of elliptical
orbital paths, or with a disk wind similar to those seen in Seyfert-1
active galactic nuclei.  We present a new analysis of the deepest
high-resolution XMM-Newton and Chandra spectra of ASASSN-14li.  Driven
by the relative strengths of He-like and H-like charge states, the
data require ${\rm [N/C]} \geq 2.4$, in qualitative agreement with UV
spectral results.  Flows of the kind seen in the X-ray spectrum of
ASASSN-14li were not clearly predicted in simulations of TDEs; this
left open the possibility that the observed absorption might be tied
to gas released in prior AGN activity.  However, the abundance pattern
revealed in this analysis points to a single star rather than a
standard AGN accretion flow comprised of myriad gas contributions.
The simplest explanation of the data is likely that a moderately
massive star ($M\gtrsim 3~M_{\odot}$) with significant CNO processing
was disrupted.   An alternative explanation is that a lower mass star
was disrupted that had previously been stripped of its envelope.  We
discuss the strengths and limitations of our analysis and these
interpretations.
\end{abstract}

\section{Introduction}
Tidal disruption events (TDE) are observed as multi-wavelength flares
when a star is disrupted by a massive black hole (see, e.g., Rees
1988).  The ready supply of gas for rapid accretion means that the
black hole can briefly accrete in excess of the Eddington limit;
this mode of accretion appears to be rare among massive black holes in
the nearby universe (e.g., Hickox et al.\ 2009).  The nature and
evolution of the accretion flow following a stellar disruption remain
a matter of debate; the key questions revolve around how quickly
strands of the disrupted star can form an accretion disk, and the
relative importance of viscous dissipation, stream collisions, and
scattering in producing the radiation that is observed.

These uncertainties are vividly illustrated by comparing the
temperatures and emitting areas of the thermal UV and X-ray continua
that are simultaneously observed in many TDEs: the temperatures are
distinct, and the emitting areas differ by an order of magnitude or
more (for reviews, see Roth et al.\ 2020, Gezari 2021).  However, when
the thermal X-ray emission from TDEs is fit with a physically
motivated model for disk emission, plausible black hole masses are
derived (e.g., Mummery \& Balbus 2020; Wen et al.\ 2020, 2023).  X-ray
observations have also revealed quasi-periodic oscillations and
relativistic reverberation suggest plausible masses (Reis et
al.\ 2012, Pasham et al.\ 2019, Kara et al.\ 2016), which is only
possible if at least some of the X-ray flux is direct emission from
the innermost accretion flow, rather than scattered light.  Optical
spectroscopy may also point to rapid circularization and disk
formation (e.g., Hung et al.\ 2020).

Narrow X-ray absorption lines have the potential to trace disk winds,
and other structures in the accretion flow that lie close to the
central engine.  Miller et al.\ (2015) detected blue-shifted
absorption lines from moderately and highly ionized charge states of
abundant elements in ASASSN-14li.  The observed blue-shifts, only
${\rm few}\times 10^{2}~{\rm km}~{\rm s}^{-1}$, are similar to those
observed in ``warm absorber'' winds in nearby Seyfert AGN.  It is
notable that the outflow speeds are in agreement with UV absorption
lines (Cenko et al.\ 2016).  The gas could represent a Seyfert-like
wind, or it could represent gas near to the apocenter of elliptical
orbits that had not yet formed into a disk.  The latter scenario would
seem to require a degree of fine-tuning, but disrupted stars may orbit
on paths that are not aligned with the angular momentum of the black
hole, potentially creating a kind of ``wicker basket'' of streams as
the TDE unfolds (see, e.g., Guillochon \& Ramirez-Ruiz 2015).

ASASSN-14li was discovered with ASASSN on 22 November 2014 (MJD 56983; Jose et al.\ 2014).  It was immediately apparent that the transient coincided with the nucleus of PGC 043234 (also known as Zw VIII 211).  At a redshift of just $z=0.0206$, or 90.3~Mpc for standard cosmological parameters, ASASSN-14li was the most proximal TDE detected in over 10 years.  Indeed, ASASSN-14li remains the clearest example of a TDE with a wind; its flux also facilitated the detection of X-ray QPOs that were stable over many weeks (Pasham et al.\ 2019), potentially indicating a misalignment of angular momentum vectors.  Although ASASSN-14li may be exceptional, it is more likely that these features are detected in this source merely owing to its high X-ray peak flux. 

The deep XMM-Newton observation wherein winds were detected is the centerpiece of the new analysis presented herein.  Prior fits to the X-ray continuum in that observation with a simple blackbody suggest a black hole mass of $M = 2.5\times 10^{6}~M_{\odot}$ via Eddington limit scaling and $M = 1.9\times 10^{6}~M_{\odot}$ via the implied emitting area (Miller et al.\ 2015).  The early decay of the the UV light curve as measured with the Swift/UVM2 filter is consistent with the $F\propto t^{-5/3}$ decay predicted by seminal theory (Rees et al.\ 1988, Phinney 1989).  Assuming this index for the early decay gives a disruption date of $t_{0} = 56948 \pm 3$.  Fits to the full multi-wavelength decay with a model including direct and reprocessed emission from an accretion disk gave a mass range of $M = 0.4-1.2\times 10^{6}~M_{\odot}$ (Miller et al.\ 2015; also see Guillochon et al.\ 2014).

In contrast to its early phase, the late evolution of ASASSN-14li between hundreds and 2600 days after the stellar disruption is consistent with a shallow decline or plateau (e.g., van Velzen 2019; Mummery \& Balbus 2020; Wen et al.\ 2020, 2023).  The late-time UV spectrum is consistent with an accretion disk, uncomplicated by an additional UV region that may be needed at early times.  Especially when multi-wavelength monitoring data are combined and analyzed within the framework of relativistic disk models, it becomes possible to constrain the total accreted mass (and the disrupted stellar mass, generally taken to be twice the accreted mass following Rees 1988), the mass and spin of the black hole, and potentially the outer radius of the accretion disk and the torque condition at the inner boundary (Mummery \& Balbus 2020; Wen et al.\ 2020, 2023).

Elemental abundances can potentially give an independent angle on the mass of the stars that are disrupted in TDEs.  Yang et al.\ (2017) present an analysis of UV spectra of ASASSN-14li, PTF15af, and iPTF16fnl, and find that ${\rm [N/C]} \geq 1.5$ is required in all cases (also see Cenko et al.\ 2016 concerning ASASSN-14li).  This is consistent with the
products of the CNO chain in moderate mass stars, enabling Yang et
al.\ (2017) to place a lower limit of $M \geq 0.6~M_{\odot}$ on the
mass of the disrupted stars in these events.  Herein, we re-examine
the most sensitive X-ray spectrum of ASASSN-14li, to test if the X-ray
data can also be used to constrain abundances and thereby the mass of
the distrupted star.  Throughout this work, we define ${\rm [N/C]}$ as the
logarithm of the ratio of the abundance of nitrogen relative to its
solar value (${A}_{N}$) and carbon relative to its solar value
(${A}_{C}$).   

Section 2 describes the observations and our data reduction.  Section
3 presents our analysis and results, using an updated photoionization
model.  In Section 4, we scrutinize our methods, explore the
implications of the inferred N/C abundance ratio, and briefly examine
the potential of future X-ray spectroscopy of TDEs.

\section{Observations and Reduction}
Miller et al.\ (2015) examined the multi-wavelength evolution of
ASASSN-14li using the XRT and UVOT telescopes aboard the Neil Gehrels
Swift Observatory.  Of the UVOT filters available, the UVM2 has the
smallest red leak, and therefore represents the best measure of the UV
flux.  The UVM2 light curve is consistent with the anticipated $F
\propto t^{-5/3}$ decay, and predicts that the disruption occurred on
MJD $56948\pm 3$.  The observations that are the focus of this paper
started on MJD 56999, approximately 50 days later.

XMM-Newton observation 0722480201 started on 08 December 2014 at
05:38:22 (UT), and obtained 95~ks.  The data from the RGS and EPIC
cameras were reduced using SAS version 1.3 (xmmsas\_20211130), and the
associated standard calibration files.  The RGS was operated in its
default ``spectroscopy'' mode.  We generated an EPIC-pn and EPIC-MOS
events lists using the tools ``epproc'' and ``emproc,'' and used the
light curves to create a filter against soft proton flaring.  We then
ran the tool ``rgsproc'' to create RGS spectral files, response files,
and background files.  In order to maximize the sensitivity within the
spectra, we restricted our analysis to the time-averaged RGS-1 and
RGS-2 spectra.

Chandra observations 17566 and 17567 started on 08 December 2014 at
23:20:28 (UT) and 11 December 2014 at 08:45:20.  Total exposures of
34.8 and 44.5~ks were obtained.  In both cases, the HRC+LETGS
combination was used to obtain high-resolution spectra.  CIAO version
4.15 and the associated CALDB files were used to create first-order
spectral files, background files, and responses from each observation.
These exposures were intended to be a single integration, and the
source varied negligibly between the exposurse, so we combined the
first-order spectra and responses using the tool
``combine\_grating\_spectra.''

\section{Analysis and Results}
Miller et al.\ (2015) analyzed X-ray spectra of ASASSN-14li using the
SPEX package (Kaastra et al.\ 1996), and reported the first direct
application of the ``pion'' photoionization model.  In this analysis,
we follow an analogous procedure using updated versions of SPEX and
``pion'' (versions 3.06.01 and 1.04, respectively).  ``Pion'' offers
two advantages over other widely available photoionization models and
packages.  First, unlike external tables of photoionization spectra
that are based on a fixed input spectrum, the ``pion'' model reads the
illuminating flux and self-consistently adjusts within the process of
miniming the goodness-of-fit statistic, so that the gas is irradiated
by the best-fit continuum.  Second, multiple ``pion'' components can
be layered -- the total spectral model can be constructed so that
radially exterior absorption zones see the central engine flux after
modification by interior absorption zones (see, e.g., Trueba et
al.\ 2019).

After obtaining spectra and responses from ``rgsproc,'' the ``trafo''
package was used to convert them into the format required by SPEX.
This conversion streamlines the files and makes spectral fits within
SPEX much faster.  We made joint fits to the RGS-1 and RGS-2 spectra
over the 16-36~\AA~ band, minimizing a Cash statistic (Cash 1979).
Outside of this pass band, the sensitivity of the spectra is greatly
reduced.  We adopted an adaptive binning scheme in order to maximize
the sensitivity of the spectra: bins in the 16--20~\AA~ range were
grouped to have a signal to noise ratio of 10.0, and bins in the
20--36~\AA~ range were grouped to have a signal to noise ratio of 5.0.
The summed first-order Chandra LETG spectrum was grouped in the same
manner and fit over the same band.  All of the errors quoted in this work are based on the values of model parameters at their $1\sigma$ confidence limits.

We initially considered the same model that is reported in Miller et
al.\ (2015): a simple blackbody (``bb'') continuum component, acted
upon by local photoionized absorption (``pion'') with solar abunances,
absorption in the hot interstellar medium of the host galaxy
(``hot''), and absorption in the ISM of the Milky Way (again via
``hot'').  The appropriate redshift was applied to all of the
components acting within ASASSN-14li and its host galaxy (via
``reds'').  The full results of this fit are listed as model ``XMMs''
in Table 1 and shown in Figure 1 and Figure 2 (in blue).

This baseline model gives a Cash statistic of $C = 1547.55$ for $\nu =
641$ degrees of freedom.  The blackbody temperature is measured to be
$kT = 50.7\pm 0.3$~eV, and its flux normalization is measured to be $K = 5.1\pm 0.4 \times
10^{25}~{\rm cm}^{2}$, or a characteristic radius of $r = 2.0\pm
0.1\times 10^{12}$~cm assuming that $K = 4\pi r^{2}$.  It is likely that these fiducial numbers are underestimates, since Compton scattering leads to artifically high
temperatures and small emitting areas (e.g., Shimura \& Takahara
1995).  The blackbody temperature is formally consistent with the
value reported in Miller et al.\ (2015), while the emitting area is
slightly higher.  The inferred luminosity of the source is $L_{X} =
2.8\pm 0.2 \times 10^{44}~{\rm erg}~{\rm s}^{-1}$ (0.1--10~keV).

The measured outflow velocity is $v = -340^{-20}_{+40}~{\rm km}~{\rm
  s}^{-1}$, with a turbulent velocity of $\sigma = 86\pm 8~{\rm
  km}~{\rm s}^{-1}$.  The outflow column density is measured to be
$N_{H} = 2.6\pm 0.3 \times 10^{22}~{\rm cm}^{-2}$, with an ionization
given by ${\rm log}\xi = 4.71\pm 0.05$.  Relative to the outflow
parameters reported in Miller et al.\ (2015), the outflow velocity is
approximately 50\% higher, the turbulent velocity is broadly
consistent, the column density is approximately two times higher, and
the ionization is about three times higher.  If the ionization
parameter is frozen to the value measured in the prior fits, ${\rm
  log}\xi = 4.2$, the fit statistic increases to $C = 1615.55$ ($\Delta C = 68$, $\Delta\nu = 1)$, strongly indicating that the change in values is real.  The
differences are likely due to improvements in the atomic physics that
is included in the ``pion'' model, and improved weighting.

Next, we considered three potential improvements to this model: (1)
non-solar values of the abundances of C and N within this single wind
zone, (2) a second, faster, more central wind zone with solar
abundances, and (3) two wind zones with non-solar but linked values
for the abundance of C and N.

When the abundances of C and N are allowed to float in the single wind
zone, the fit statistic improves to $C = 1351.2$ for $\nu = 639$
degrees of freedom (or $\Delta C = -196.28$ for $\Delta\nu = -2$; see model ``XMMt'' in Table 1, and model in red in Figure 1 and Figure 2 and the discussion below).  When the abundances of C and N are fixed at solar values
but a faster, radially interior wind zone is added, the fit statistic
improves to $C = 1490.41$ for $\nu = 639$ degrees of freedom (or
$\Delta C = -57.14$ for $\Delta\nu = -2$).  In this fit, the velocity
width of the fast absorber was constrained to be 10\% of its outflow
velocity, in order to keep the fast zone from locking onto narrow
lines that might be better associated with slower gas, or very broad
features that might not be tied to the flow.  The nominal outflow
velocity is $v = -24,000\pm 1000~{\rm km}~{\rm s}^{-1}$.  The best
overall fit is achieved when two wind zones are included, with linked
but variable values for the abundance of C and N.  This model achieves
$C = 1343.48$ for $\nu = 637$ ($\Delta C = -204.07$ for $\Delta \nu = -4$).

Most of the improvement in the this model comes from allowing C and N
to have non-solar abundances, and evidence for a very fast wind is
marginal.  Applying the Akaike Information Criterion as implemented by
Emmanoulopoulos et al.\ (2016) and judging the improvements in terms
of $\Delta$AIC, the abundance variations are highly significant
whereas the addition a faster outflow is not statistically significant
($\Delta$AIC $<$ 2).  In all subsequent fits, and in all fits listed
in Table 1, we focused on single-zone wind models and the role of
abundances.

In model ''XMMf,'' the abundances of carbon and nitrogen were fixed
to arbitrary but plausible values (based on Mockler et al.\ 2022), in
order to make an initial examination of the sensitivity of the data to
changes in abundance.  We set ${\rm A_{C}} = 0.3$ and ${\rm A_{N}} = 10$.  This
model produced a statistically significant improvement over ``XMMs''
with solar abundances for all elements: $C = 1488.12$ for $\nu = 641$
degrees of freedom ($\Delta C = -59.43$ for $\Delta \nu = 0$).  We next ``thawed'' the abundances of carbon and
nitrogen to vary freely within the fit (formalizing the exploratory
fit made previously).  Model ``XMMt'' finds a carbon abundance of
${\rm A_{C}} = 0.0^{+0.1}$, and a nitrogen abundance of ${\rm A_{N}} = 110\pm20$
(1$\sigma$ errors.  This model represents another statistically
significant improvement: $C = 1351.2$ for $\nu = 639$ degrees of
freedom ($\Delta C = -196.35$ for $\Delta \nu = -2$).

These results are formally consistent with a carbon abundance of zero.  This reflects a limitation of the data, not a physical reality.  In far more sensitive data, it is likely that the H-like C line would be detected, and a formal lower limit would be obtained.  While a formal measurement of [N/C] with attendant errors is not possible with these data, we can report a lower limit.  Taking the lower limit for nitrogen abundance relative to solar, ${\rm A}_{N} = 90$, and the upper limit for carbon, ${\rm A}_{\rm C} = 0.1$, we obtain a lower limit of [N/C] $\geq$ 3.0.  Given the extremity of this limit, we next undertook a set of additional fits to understand if the inferred abundance pattern is influenced by any instrumental issues, and report limits on [N/C] in the same manner.

We first decoupled the abundances of carbon and nitrogen between the RGS1 and RGS2 spectra.  This modification tests the possibility that the extreme abundances are driven by an anomaly in one or both instruments that coincides with the key lines.  The ``XMMd'' column in Table 1 lists the abundances derived from each instrument.  The upper limit on the abundance of carbon derived in the RGS2 spectrum is four times higher than that derived in the RGS1 spectrum (${\rm A_{C}} \leq 0.4$ versus ${\rm A_{C}} \leq 0.1$).  The constraints derived on the nitrogen abundance in each spectrum are closely consistent.  If we conservatively consider the less constraining limit on ${\rm A}_{\rm C}$, our results indicate ${\rm [N/C]} \geq 2.4$.  The $3\sigma$ upper limits on the carbon abundances for RGS 1 and RGS2 are ${\rm A_{C}}\leq 1.3$ and ${\rm A_{C}} \leq 1.6$, respectively.  For both RGS1 and RGS2, the measured nitrogen abundances and associated $3\sigma$ errors are
${\rm A_{N}} = 110^{+60}_{-40}$.  In short, solar abundances of carbon
are allowed within $3\sigma$, but much higher abundances of nitrogen
are still required.

Fits to the Chandra spectrum are listed in columns ``CXOs'' and
``CXOt'' in Table 1, designated to correspond to the fits made to the
XMM-Newton spectra with ``solar'' and ``thawed'' abundances.  The
Chandra spectrum is less sensitive than the XMM-Newton spectra, but
these fits serve as a useful check since the instrument profile is
independent (including background, calibration, systematic errors, etc.).
Again, the fit with variable abundances is a significant improvement
over the fit with solar abundances ($C/\nu = 328.0/278$ versus $C/\nu
= 351.5/280$, or $\Delta C = -23.5$ for $\Delta \nu = -2$).  The upper limit on the C abundance is ${\rm A_{C}} \leq
0.4$, and the N abundance is measured to be ${\rm A_{N}} =
160^{+210}_{-60}$.  Nominally, then, ${\rm [N/C]} \geq 2.4$.  Given that
independent instruments both prefer a sub-solar C abundance and
elevated N abundance, and also given the similarity of the
constraints, this extreme ratio limit may be robust.  We additionally note that extending the Chandra LETG spectrum to 60$\AA$ does not alter the ${\rm [N/C]}$ ratio limit that is inferred.

We also examined whether or not the assumed continuum model influences the abundance ratio limit.  The true continuum flux may be that of an accretion disk, potentially modified by relativistic effects and Comptonization (e.g., Mummery \& Balbus 2020; Wen et al.\ 2020, 2023).  Over a narrow pass band and at modest sensitivity, these effects cannot be detected.  However, it is worth investigating if the slightly different continuum produced by a multi-temperature disk model provides any improvement in the Cash statistic, or in the derived abundances.  To this end, we replaced the simple blackbody in models ``XMMf'' and ``XMM'' in Table 1, with the ``dbb'' model within SPEX.  This model is based on a Shakura-Sunyaev disk (Shakura \& Sunyaev 1973); it includes a torque-free inner boundary condition and its free parameters include the characteristic disk temperature and emitting area.  

Fits with ``dbb'' produce slightly worse fits than the simple blackbody when the C and N abundances are frozen at solar values: $C = 1569.1$ for $\nu = 642$ degrees of freedom, whereas the simple blackbody achieved $C = 1547.6$ for $\nu = 641$.  This model gives a disk temperature of $kT = 0.12\pm 0.01$~keV and an emitting area of $K = 1.8\times 10^{24}~{\rm cm}^{2}$.  When ``dbb'' replaces the simple blackbody within model ``XMMt,'' wherein the abundances of C and N are allowed to vary, a statistically equivalent fit is achieved.  Importantly, however, the same abundance limit and measurement result: ${\rm A}_{\rm C} = 0.0^{+0.1}$ and ${\rm A}_{\rm N} = 120^{+16}_{-14}$.  This strongly indicates that the derived abundance ratio limit does not depend on the specific continuum model assumed.

Finally, we note that none of the fits detailed in this section are formally acceptable.  Despite this, changes in the fit statistic owing to model changes and enhancements are meaningful, and direct comparisons yield physical insights.  The absence of a statistically acceptable model is typical of line-rich high-resolution X-ray spectra, wherein even sophisticated models necessarily offer an imperfect description of a complex physical scenario.  For instance, in recent fits to RGS spectra from the Seyfert-1 AGN 3783, Mao et al.\ (2019) only achieve a fit statistic of $C = 6092$ for $\nu = 2505$ degrees of freedom, despite many layers of ionized absorption and a combination of broad and narrow emission lines.

\section{Discussion}
We have re-analyzed the most sensitive high-resolution X-ray spectra
obtained from ASASSN-14li, with a novel focus on key elemental
abundances.  The spectra can be described using two zones with a wide
separation in velocity, but a single zone with non-solar abundances is
strongly preferred.  We find that a model that assumes solar
abundances for all elements under-fits He-like and H-like N absorption
lines, and over-predicts the strength of the H-like C line (see Table
1, Figure 1, and Figure 2).  Allowing for potential differences
between the RGS1 and RGS2 spectra in the region of the H-like C line,
the XMM-Newton data still require ${\rm [N/C]} \geq 2.4$; the Chandra
spectrum of ASASSN-14li independently verifies this ratio limit.  The same limit is obtained when a disk blackbody continuum is assumed instead of a simple blackbody.  Here, we discuss the implications of this finding for our understanding of TDEs.

The CNO cycle operating in stellar interiors leads to core material that is nitrogen-rich and carbon-deficient relative to material unprocessed by nuclear burning (Iben 1964, Iben 1967,
Lambert \& Ries 1977, Lambert \& Ries 1981).  The rate of CNO
processing is regulated by the Coulomb barrier for proton capture onto
CNO nuclei and consequently has a strong temperature
dependence.  Increasing core temperature with stellar mass means that
CNO processing is more efficient and more extensive in higher mass
stars, leading to greater N enrichment and C depletion (Iben 1964).

Kochanek et al.\ (2016) and Gallegos-Garcia et al.\ (2018) explored
the potential for this processing to reveal the masses of disrupted
stars.  Mockler et al.\ (2022) recently studied this issue in detail,
finding that stars more massive than $M = 1.3~M_{\odot}$ are required
to give ${\rm [N/C]} \geq 1.5$.  Based on stellar demographics, the
fact that massive stars live for a relatively short time, and the
properties of stars observed in nuclear star clusters, Mockler et
al.\ (2022) conclude that stars more massive than $M = 3~M_{\odot}$
are not likely to be disrupted at an significant rate.

Fits to the UV spectrum of ASASSN-14li require ${\rm [N/C]} \geq 1.5$
(Yang et al.\ 2017).  The more extreme abundance ratio in our fits to
the X-ray data of ASSASN-14li nominally point to a star at the limit
of the range considered by Mockler et al.\ (2022), or potentially an
even more massive star.  If the cluster of young, massive stars
surrounding Sgr A* (e.g., Lu et al.\ 2013) is typical of the nuclear
environment in other galaxies, then it is possible that a sizable
fraction of disruptions involve relatively massive stars.

Particularly if the diffuse UV and X-ray gas have different abundance
patterns -- potentially corresponding to different parts of the
stellar interior -- their relative position within the accretion flow
may be a window on the disruption itself.  However, such disparities
are not anticipated by theoretical treatments that address mixing
(e.g., Law-Smith et al.\ 2019), and we note that the narrow UV and
X-ray absorption lines in ASASSN-14li have fully consistent
blue-shifts (Miller et al.\ 2015, Cenko et al.\ 2016).  The broadening
of the N IV] emission line in the Hubble UV spectrum of ASASSN-14li
  suggests a production radius of about 150~AU, or $r \simeq 2.3\times
  10^{15}~{\rm cm}$ (Cenko et al.\ 2016).  The gas exhibiting H-like
  and He-like N absorption in X-rays is constrained to lie within $r
  \leq 3\times 10^{15}~{\rm cm}^{-2}$ through variability (Miller et
  al.\ 2015).

Even with realistic mixing between layers at disruption, a full
disruption of a $M \simeq 3~M_{\odot}$ star may not be able to
reproduce the abundances measured in this study (see Law-Smith et
al.\ 2019).  It is possible that the environement close to massive
black holes can lead to mergers, via the ``eccentric Kozai Lidov''
mechanism (Stephan et al.\ 2016).  Such stars could resemble G2 and
represent a young, massive population that is not possible in less
extreme environments.  But, more mundane explanations are also
possible.  Even within solar-mass stars, nitrogen is more abundant
than carbon close to the core.  If the disrupted star had previously
been partially stripped of its envelope, exposing more processed
material at the point of disruption, this could potentially account
for at least some of the observed abundance pattern.  Key facets of
the nuclear environment could also play a role: McKernan et
al.\ (2022) note that the envelopes of evolved stars can be stripped
by ram pressure in AGN disks.  A potential shortcoming of this
alternative is that a strong Ly$\alpha$ line was detected in
ASASSN-14li (Cenko et al.\ 2015), signaling that at least some of the
stellar envelope was retained prior to the disruption event.  If
stripping is not viable as a fully independent alternative to the
massive stellar interpretation, it may be that a degree of stripping
contributed to the extreme abundance pattern.

Theoretical treatments of TDEs did not anticipate the discovery of
X-ray winds with properties like Seyfert warm absorbers.  In the
classical picture of TDEs, approximately half of the stellar mass is
expected to be ejected at much greater speeds, $v \geq 10^{4}~{\rm
  km}~{\rm s}^{-1}$ (e.g., Rees 1988).  Variability in the wind
observed in ASASSN-14li pointed to an absorption radius consistent
with the broad line region (BLR) in Seyferts (Miller et al.\ 2015),
and it is notable that the best optical spectra of TDEs now also
exhibit a BLR (e.g., Hung et al.\ 2020).  In this sense, UV and X-ray
winds consistent with Seyfert BLRs might be expected to be common in
TDEs.  However, it was not possible to completely exclude the
possibility that the X-ray wind could be tied to {\it prior} low-level
AGN activity within the nucleus.

Studies that utilize optical emission lines from the BLR in AGN
generally find super-solar metallicities (e.g., Ferland et al.\ 1996,
Dietrich et al.\ 1999, 2003).  However, several effects can complicate
such inferences.  In principle, at least, wind absorption lines offer
the chance to make absolute measurements with direct ratios to
hydrogen.  Combining simultaneous UV and X-ray wind spectra of Mrk
279, for instance, Arav et al.\ (2007) found that the abundances of C,
N, and O are elevated by factors of $2.2\pm 0.7$, $3.5\pm 1.1$, and
$1.6\pm 0.8$ (respectively).  This pattern differs markedly from the
extreme nitrogen to carbon ratio limit inferred in ASASSN-14li.  As a
result, the wind observed in ASASSN-14li can only be consistent with
material that originated in a single, disrupted star.

Mummery \& Balbus (2020) fit the X-ray and UV decay curves of ASASSN-14li with a relativistic thin disk model, and find that the accreted mass was likely $M \sim 0.016~M_{\odot}$, implying a disrupted stellar mass of $M \sim 0.032~M_{\odot}$.  In contrast, Wen et al.\ (2020) fit the X-ray decay with slim accretion disk models that include advection and a spectral hardening factor, and find that a disrupted stellar mass of $M \geq 0.34~M_{\odot}$ is implied.  It is possible that these low disrupted stellar mass estimates can be reconciled with the relatively high mass inferred via abundances if the accretion disk formation efficiency is very low (see, e.g., Dai et al.\ 2015,  Piran et al.\ 2015, Shiokawa et al.\ 2015), so that most of the disrupted stellar mass is lost and/or waiting to accrete even at very late times.  However, both treatments omit the role of outflows in assessing the total accreted mass, though both winds and jets are clearly present in ASASSN-14li (e.g., Miller et al.\ 2015, Alexander et al.\ 2016).  A slim disk with strong advection and correspondingly lower efficiency for converting ${\dot M}$ into radiation would also lead to higher estimates of the accreted mass and disrupted stellar mass.  Additionally, Wen et al.\ (2023) note that if the UV emission observed in ASASSN-14li soon after its discovery is due to shocks, a disrupted stellar mass in excess of $M \geq 1.4~M_{\odot}$ is required for their minimum preferred black hole mass.  Most importantly, perhaps, modeling of the UV light curve of ASASSN-14li points to a disruption about $35\pm3$ days before its discovery, during a time when it was unobservable from the ground (Miller et al.\ 2015, Holoien et al.\ 2016).  This period would likely correspond to the most highly super-Eddington part of the TDE, and would add greatly to the accreted mass and the corresponding disrupted stellar mass.  

The tools and techniques that have permitted discoveries in
ASASSN-14li are essentially those developed in the study of the
brightest Seyfert AGN.  However, the Seyfert-like flux of ASASSN-14li
is the result of its proximity.  It is now clear that the rate of
highly proximal TDEs is relatively low, so few cases will reach a
Seyfert-like flux of $F = 1.0\times 10^{-11}~{\rm erg}~{\rm
  cm}^{-2}~{\rm s}^{-1}$ in soft X-rays.  A much larger set of TDEs
are at least an order of magnitude fainter.  Given that long
observations are needed to perform detailed spectroscopic and
variability studies in bright TDEs and Seyferts using Chandra,
XMM-Newton, and NuSTAR, it is clear that additional progress will
require more sensitive instruments.  The Athena/X-IFU will provide a
resolution of 2~eV across the 0.3-10.0~keV pass band, and an effective
area nearly two orders of magnitude higher than the RGS in the
0.3--0.5~keV band where the crucial N and C lines are found (Barret et
al.\ 2022).  The proposed Arcus Probe mission will deliver even higher
resolution at long wavelengths, and simultaneous UV coverage (Smith et
al.\ 2023).

We thank the anonymous referee for comments that improved this manuscript.
We acknowledge helpful discussions with Elena Gallo, Jelle Kaastra,
and Jelle de Plaa.  We note that the ``wicker basket'' scenario
referenced in this work may have been coined by Martin Rees at a
conference in his honor at the University of Cambridge in 2017.

\clearpage

\begin{table}[h]
\caption{Spectral models and parameters}
\begin{footnotesize}
\begin{center}
\begin{tabular}{lllllll}    
Parameter          &                                    XMMs  &    XMMf  &    XMMt &  XMMd &  CXOs  & CXOt \\
\tableline
abundances     &                                        solar    &  frozen     &    thawed   &  thawed, dcpl. &   solar &  thawed \\
\tableline
kT (eV) &                                               $50.7\pm 0.3$  & $51.4\pm 0.4$ & $54.6\pm0.5$ & $54.6\pm 0.5$ & $53.2\pm 0.9$ & $55.9\pm0.9$  \\ 
Norm $(10^{25}~{\rm cm}^{2}$) &                           $5.1\pm 0.4$  & $4.4\pm 0.4 $ & $2.2\pm 0.2$ & $2.2\pm0.2$ & $2.7\pm0.6 $  & $1.7\pm 0.4$  \\ 
\tableline
$N_{\rm H}$ (TDE, $10^{22}~{\rm cm}^{-2}$) &                 $2.6\pm 0.3$ & $3.0^{+0.5}_{-0.4}$ & $2.5\pm 0.4$ & $2.5\pm0.4$ & $1.8\pm0.6$ & $2.0^{+0.9}_{-0.6} $  \\ 
log$\xi$ &                                               $4.71\pm0.05$  & $4.55\pm 0.07$ & $4.32\pm 0.06$ & $4.31\pm0.04$ & $4.6\pm0.1 $ & $4.4\pm0.1$\\ 
v$_{rms}$ (${\rm km}~{\rm s}^{-1}$) &                      $86\pm 8$   & $76\pm 7$ & $91\pm 8$ & $91\pm8$ & $67\pm14$ & $60\pm 12$ \\ 
v$_{out}$ (${\rm km}~{\rm s}^{-1}$) &                      $340\pm 30$  & $360^{+20}_{-30}$ & $370\pm 30$ & $370\pm30$ & $280\pm80$ & $330^{+70}_{-50}$  \\ 
A$_{\rm C}$ &                                                1.0*  & 0.33* & $0.0^{+0.1}$ & $0.0^{0.1}$, $0.0^{+0.4}$ & 1.0* & $0.0^{+0.4}$    \\ 
A$_{\rm N}$ &                                                1.0*  & 10.0* & $110\pm20$ & $113^{+17}_{-15}$, $109^{+17}_{-15}$ & 1.0* & $160^{+210}_{-60}$ \\ 
\tableline
$N_{\rm H}$ (host, $10^{20}~{\rm cm}^{-2}$) &               $1.1^{+0.6}_{-0.3}$  & $1.3^{+0.5}_{0.4}$ & $1.2\pm 0.4$ & $1.2\pm0.4$ & $2.0\pm 0.9$ & $1.4\pm0.7$ \\ 
\tableline
$N_{\rm H}$ (MW, $10^{20}~{\rm cm}^{-2}$) &                 $4.3^{+0.4}_{-0.7}$ & $3.8^{+0.4}_{-0.5}$ & $2.8\pm 0.4$ & $2.1\pm 0.4$ & $2.5\pm 0.8$ & $1.9\pm0.7$\\  
\tableline
Flux $(10^{-11}~{\rm erg}~{\rm cm}^{-2}~{\rm s}^{-1}$) &   $3.0\pm 0.2$ & $3.1\pm 0.3$ & $3.5\pm 0.4$ & $3.6\pm0.4$ & $2.8\pm0.6$  & $3.2\pm0.8$ \\  
\tableline
C/${\nu}$ &                                             1547.6/641  & 1488.12/641 & 1351.2/639  & 1351.1/637 & 351.5/280 & 328.0/278 \\ 
\tableline
\end{tabular}
\vspace*{\baselineskip}~\\
\end{center} 
\tablecomments{Parameters measured from models fit to the
  XMM-Newton and Chandra spectra of ASASSN-14li.  From top to bottom,
  the model parameters are arranged to reflect the light path taken
  from the central engine to the telescope.  Ionized absorption within
  the central engine of the TDE was fit using the ''pion''
  photoionization package; intervening absorption in the hot ISM of
  the host and Milky Way were fit using ``hot.''  Within the models,
  the blackbody emission, ionized absorption, and ISM absorption
  within the host were redshifted to the host value, so outflow
  velocities are reported relative to the host.  Parameter values that
  are marked with an asterisk (*) were fixed within the particular
  model.  The fit designated ``XMMs'' represents a joint fit to the
  RGS1 and RGS2 spectra with ``pion'' abundances fixed at solar
  values.  The fit designated ``XMMf'' froze the abundances of carbon
  and nitrogen within ``pion'' to arbitrarily diminished and enhaced
  values.  In ``XMMt,'' the carbon and nitrogen abundances were
  ``thawed'' and varied freely.  The fit designated ``XMMd'' allowed
  the abundances of carbon and nitrogen to float freely and decoupled
  these parameters in the RGS1 and RGS2 spectra.  In the ``CXOs'' fit,
  the Chandra data were fit assuming a wind with solar abundances.
  Finally in the ``CXOt'' fit, the Chandra data were fit with the
  abundances of C and N ``thawed.''  All fluxes are
  quoted in the 0.1--10.0~keV band.}
\end{footnotesize}
\end{table}
\medskip

\clearpage

\begin{figure}[t!]
    \centering
    \includegraphics[width=1.0\textwidth]{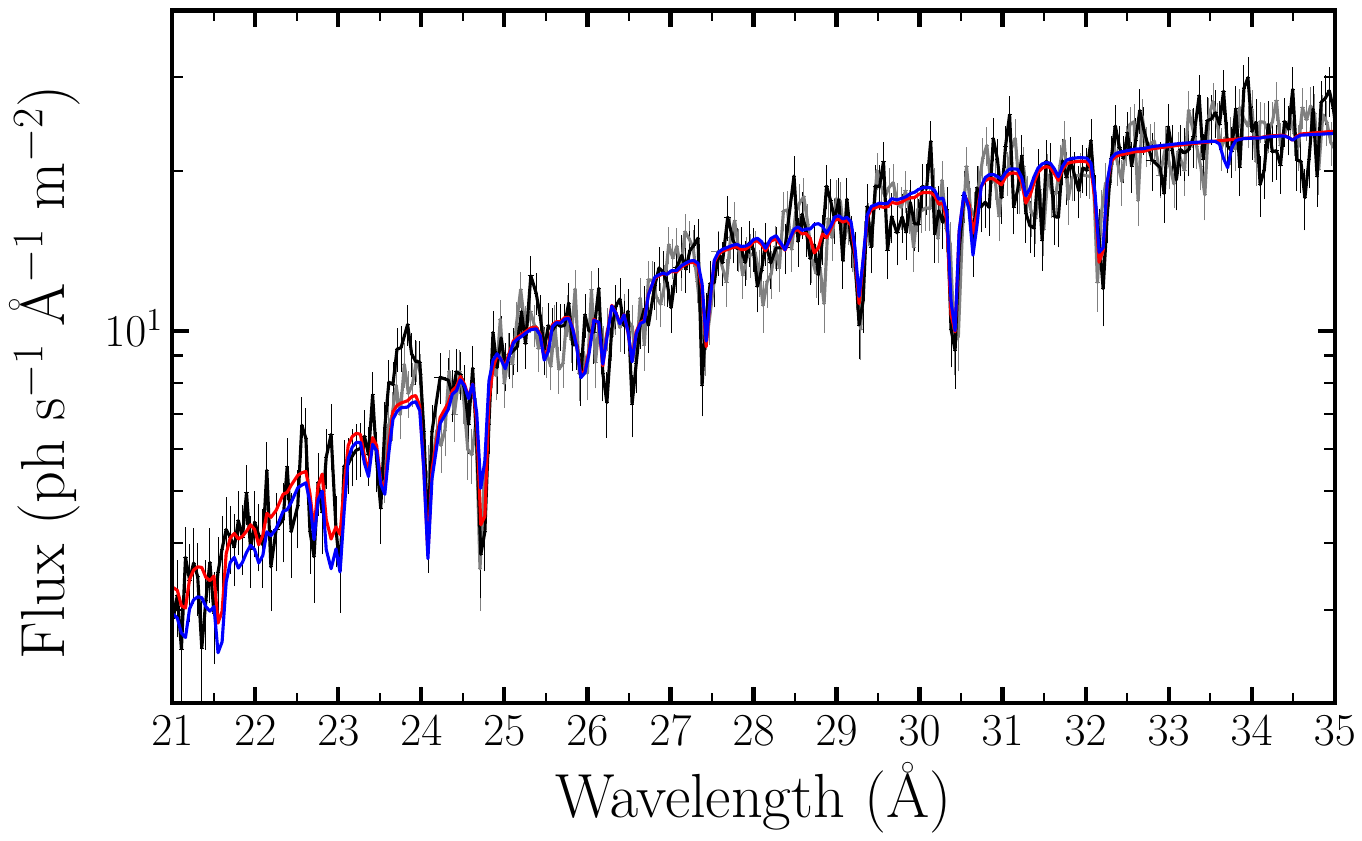}
    \vspace{-1.0in}
    \figcaption[t!]{RGS1 (black) and RGS2 (gray) spectra of
  ASASSN-14li from the longest XMM-Newton observation of this TDE.
  The wavelength scale is shifted to the frame of the host galaxy.
  The model in blue (XMMs in Table 1) represents the best ``pion''
  photoionized absorption model with solar abundances.  The H-like N
  VII resonance line at 24.78~\AA~ and the He-like N VI resonance line
  at 28.78~\AA~ are under-predicted, while the H-like C VI resonance
  line at 33.73~\AA~ is over-predicted.  The model in red (XMMt in
  Table 1) shows the best absorption model when the abundances of
  carbon and nitrogen are allowed to vary.  An enhanced N abundance,
  potentially as high as 100 times solar, provides a better fit to the
  N lines.  Similarly, the data require that the C abundance is at
  most 0.4 times solar, giving ${\rm [N/C]} \geq 2.4$.
  Please see the text and Table 1 for additional details. }
    \label{fig:newerflux}
\end{figure}

\clearpage

\begin{figure}
  \includegraphics[width=0.425\textwidth]{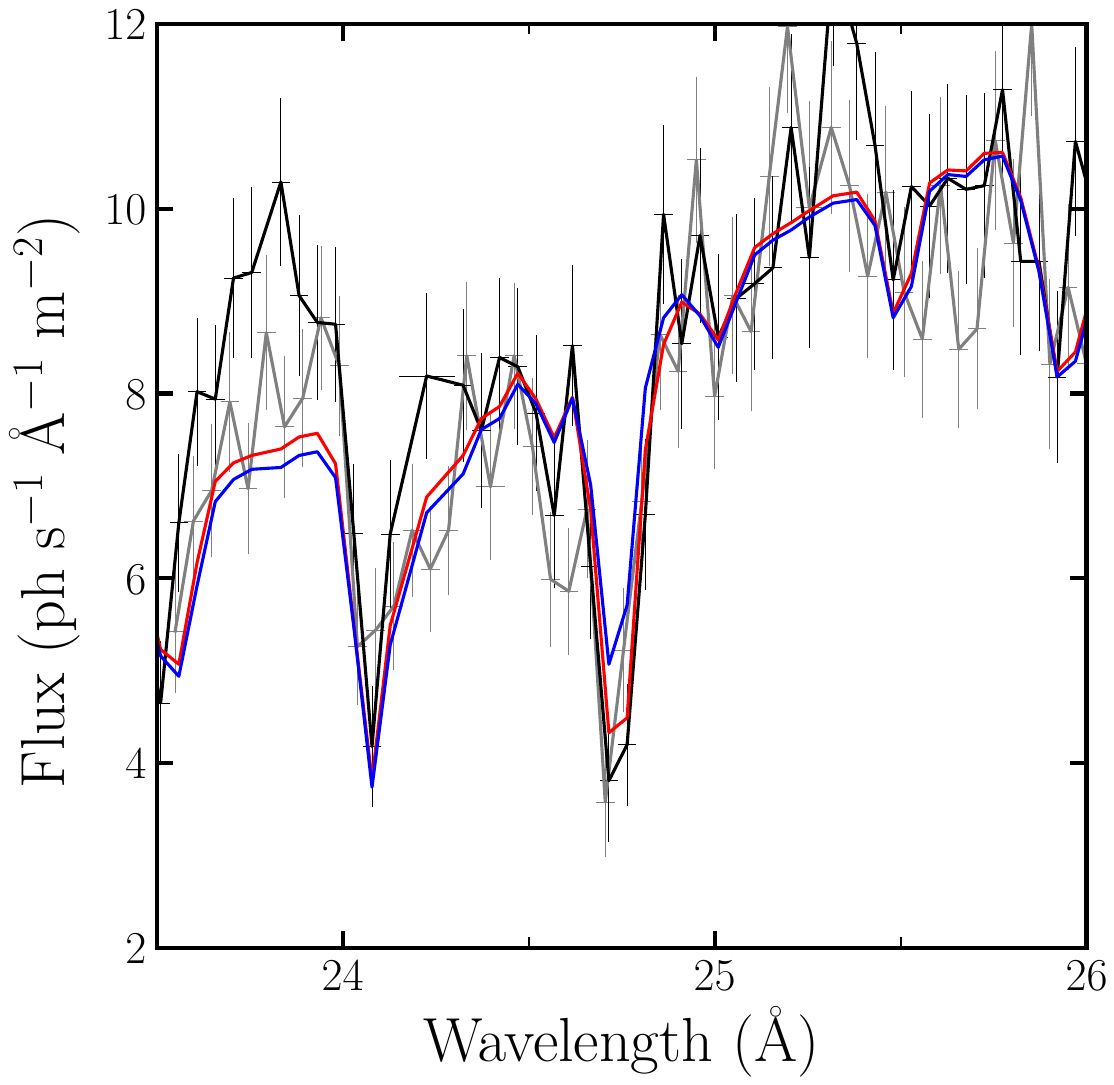}
  \hspace{-1.1in}
  \includegraphics[width=0.425\textwidth]{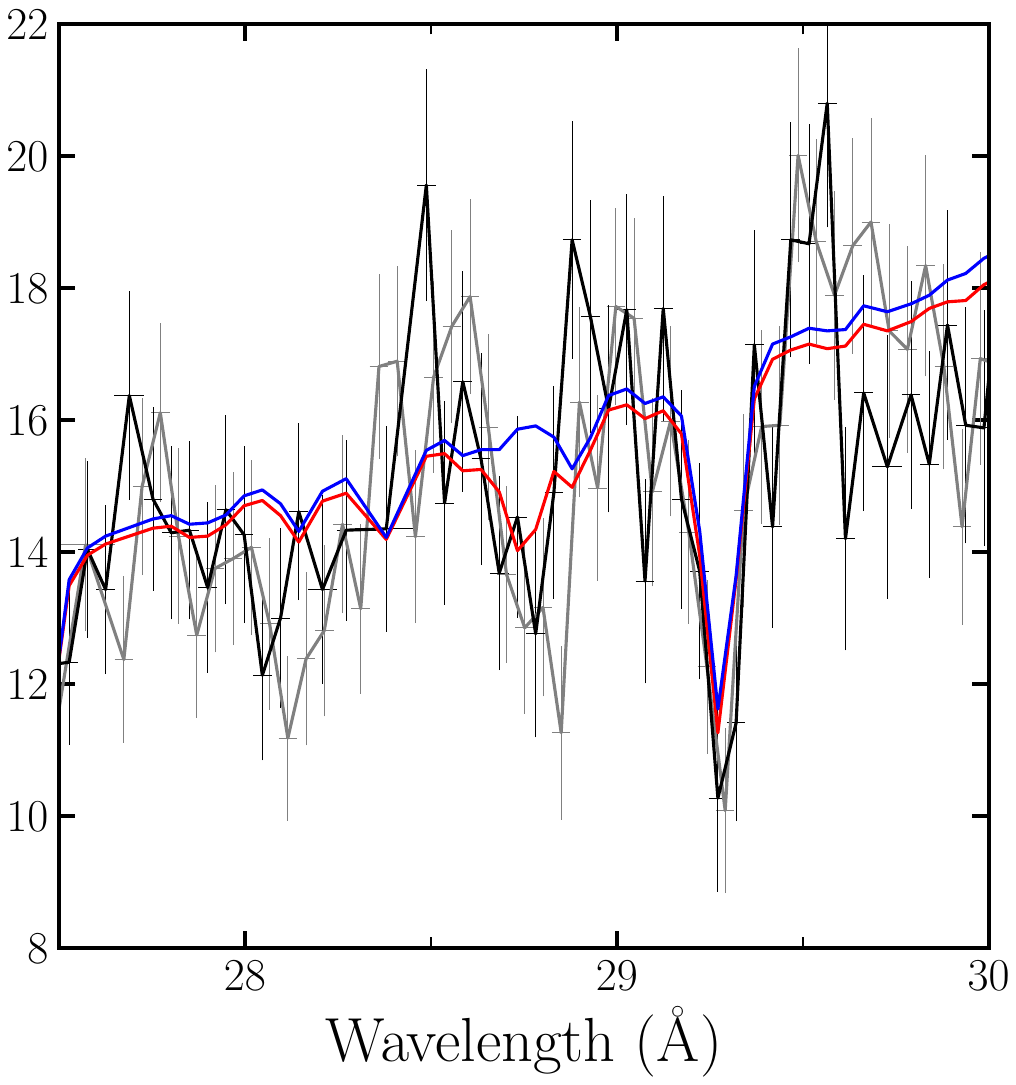}
  \hspace{-1.1in}
  \includegraphics[width=0.425\textwidth]{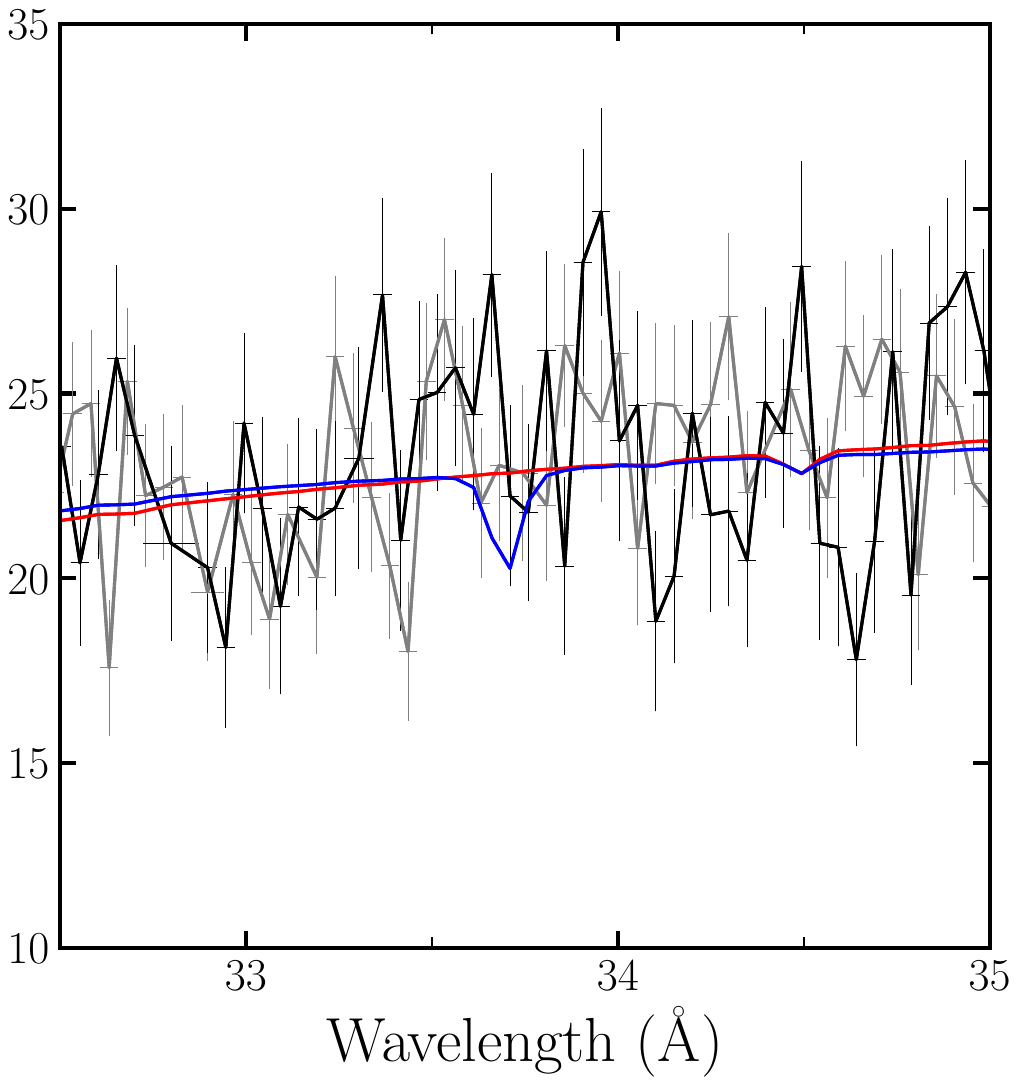}
\figcaption[t]{Narrow, 2.5\AA~ slices of the XMM-Newton
  spectra of ASASSN-14li shown in Figure 1.  The RGS1 spectrum is
  shown in black, the RGS2 spectrum in gray.  Both spectra are shifted
  to the host frame.  The model in blue is ``XMMs'' with solar
  abundances; the model in red is ``XMMt'' with thawed N and C
  abundances, giving ${\rm [N/C]} \geq 2.4$.  The left panel centers the
  H-like N VII line at 24.78\AA, the middle panel centers the He-like
  N VI line at 28.78\AA, and the right panel centers the H-like C VI
  line at 33.73\AA.}
\vspace{0.25in}
\end{figure}
\medskip

\medskip


\begin{references}
\reference{} Alexander, K., Berger, E., Guillochon, J., Zauderer, B. A., \& Williams, P. K. G., 2016, ApJ, 819, L25
\reference{} Arav, N., Gabel, J., Korista, K., Kaastra, J., Kriss, G.,
et al., 2007, ApJ, 658, 829
\reference{} Barret, D., Albouys, V., den Herder, J.-W., Piro, L.,
Cappi, M., et al., 2022, Experimental Astronomy, in press,
arxiv:2208.14562
\reference{} Cash, W., 1979, ApJ, 228, 939
\reference{} Cenko, S. B., Cucchiara, A., Roth, N., Veilleux, S.,
Prochaska, J., et al., 2016, 818, L32
\reference{} Dai, L., McKinney, J.  C., \& Miller, M. C., 2015, ApJL, 812, L39
\reference{} Dietrich, M., Appenzeller, I., Hamann, F., Heidt, J.,
Jager, K., Vestergard, M., Wagner, S. J., 2003, A\&A, 398, 891
\reference{} Dietrich, M., Wagner, S., J., Courvoisier, T., Bock, H.,
North, P., 1999, A\&A, 351, 31
\reference{} Ferland, G., Baldwin, J., Korista, K., Hamann, F.,
Carswell, R., Phillips, M., Wilkes, B., \& Williams, R., 1996, ApJ,
461, 683
\reference{} Gallegos-Garcia, M., Law-Smith, J., \& Ramirez-Ruiz, E.,
2018, 857, 109
\reference{} Gezari, S., 2021, ARA\&A, 59, 21
\reference{} Guillochon, J., Makukian, H., \& Ramirez-Ruiz, E., 2014,
ApJ, 783, 23
\reference{} Guillochon, J., \& Ramirez-Ruiz, E., 2015, ApJ, 809, 166
\reference{} Hickox, R., Jones, C., Forman, W., Murray, S., Kochanek,
C., et al., 2009, ApJ, 686, 891
\reference{} Holoien, T., Kochanek, C., Prieto, J., Stanek, K., Dong,
S., et al., 2016, MNRAS, 455, 2918
\reference{} Hung, T., Foley, R., Ramirez-Ruiz, E., Dai, L., Auchettl,
K., et al., 2020, ApJ, 903, 31
\reference{} Iben, I., 1964, ApJ, 140, 1631
\reference{} Iben, I., 1967, ApJ, 147 624
\reference{} Jonker, P., 
\reference{} Jose, J., Guo, Z., Long, F., Herczeg, G., Dong, S., et
al., 2014, The Astronomer's Telegram, 6777
\reference{} Kaastra, J., Mewe, R., \& Nieuwenuijzen, H., 1996, in
``UV and X-ray Spectroscopy of Astrophysical Plasmas,''
eds. K. Yamashitea \&T. Watanabe, 411-414
\reference{} Kara, E., Miller, J. M., Reynolds, C. S., Dai, L., 2016,
Nature, 535, 388
\reference{} Kochanek, C., et al., 2016, ApJ, 458, 127
\reference{} Lambert, D., \& Ries, L., 1977, ApJ, 217, 508
\reference{} Lambert, D., \& Ries, L., 1981, 248, 228
\reference{} Law-Smith, J., Guillochon, J., \& Ramirez-Ruiz, E., 2019,
ApJ, 882, L25
\reference{} Lu, J. R., Do, T., Ghez, A. M., Morris, M. R., Yelda, S.,
Matthews, K., 2013, ApJ, 764, 155
\reference{} Mao, J., Mehdipour, M., Kaastra, J. S., Costantini, E., Pinto, C., et al., 2019, A\&A, 621, 99
\reference{} Miller, J. M., Kaastra, J., Miller, M., Reynolds, M. T.,
Brown, G., et al., 2015, Nature, 526, 542
\reference{} Mockler, B., Guillochon, J., Ramirez-Ruiz, E., 2019, ApJ,
872, 151
\reference{} Mockler, B., Twum, A., Auchettl, K., Dodd, S., French,
K., Law-Smith J., Ramirez-Ruiz, E., 2022, ApJ, 924, 70
\reference{} Mummery, A. \& Balbus, S. A., 2020, MNRAS, 492, 5655
\reference{} Pasham, D., Remillard, R., Fragile, P., Franchini, A.,
Stone, N., et al., 2019, Science, 363, 531
\reference{} Phinney, E. S., 1989, ``Manifestations of a Massive Black Hole in the Galactic Center,'' in Morris, M. (ed.) The Center of the Galaxy, vol. 136 of IAU Symposium, 543
\reference{} Piran, T., Svirski, G., Krolik, J., Cheng, R. M., \& Shiokawa, H., 2015, ApJ, 806, 164
\reference{} Rees, M., et al., 1988, Nature, 333, 523
\reference{} Reis, R., Miller, J. M., Reynolds, M. T., Gultekin, K.,
Maitra, D., King, A. L., Strohmayer, T. E., 2012, Science, 337, 949
\reference{} Roth, N., Rossi, E. M., Krolik, J., Piran, T., Mockler,
B., Kasen, D., 2020, Space Science Reviews, 216, 114
\reference{} Shakura, N. \& Sunyaev,  R., 1973, A\&A, 24, 337
\reference{} Shimura, T., \& Takahara, F., 1995, ApJ, 445, 780
\reference{} Shiokawa, H., Krolik, J. H., Cheng, R. M., Piran, T., \& Noble, S. C., 2015, ApJ, 804, 85
\reference{} Smith, R., et al., 2023, SPIE, in preparation
\reference{} Stephan, A., Naoz, S., Ghez, A., Witzel, G., Sitarski,
B., Do, T., Kocsis, B., 2016, MNRAS, 460, 3494
\reference{} Trueba, N., Miller, J. M., Kaastra, J., Zoghbi, A.,
Fabian, A. C., Kallman, T., Proga, D., Raymond, J., 2019, ApJ, 886,
104
\reference{} Wen, S., Jonker, P. G., Stone, N. C., Zabludoff, A. I., \& Psaltis, D., 2020, ApJ, 897, 80
\reference{} Wen, S., Jonker, P. G., Stone, N. C., van Velzen, S., \& Zabludoff, A. I., 2023, MNRAS, in press, arxiv:2304.00428
\reference{} Yang, C., Wang, T., Ferland, G., Dou, L., Zhou, H.,
Jiang, N., Sheng, Z., 2017, ApJ, 846, 150

\end{references}
\end{document}